\documentclass[reprint,superscriptaddress,onecolumn,showkeys,amsmath,amssymb,aps,prf]{revtex4-2}
\usepackage{graphicx}
\usepackage{amsmath}
\usepackage{dcolumn}
\usepackage{bm}
\usepackage[dvipsnames]{xcolor}
\usepackage{caption}
\usepackage{subcaption}
\begin{document}
\title{Capillary-lubrication force between rotating cylinders separated by a fluid interface}
\author{Aditya Jha}
\affiliation{Univ. Bordeaux, CNRS, LOMA, UMR 5798, F-33405 Talence, France.}
\author{Yacine Amarouchene}
\affiliation{Univ. Bordeaux, CNRS, LOMA, UMR 5798, F-33405 Talence, France.}
\author{Thomas Salez}
 \email{thomas.salez@cnrs.fr}
\affiliation{Univ. Bordeaux, CNRS, LOMA, UMR 5798, F-33405 Talence, France.}
\date{\today}
\begin{abstract}
Two cylinders rotating next to each other generate a large hydrodynamic force if the intermediate space is filled with a viscous fluid. Herein, we explore the case where the cylinders are separated by two layers of viscous immiscible fluids, in the limit of small capillary deformation of the fluid interface. As the interface deformation breaks the system's symmetry, a novel force characteristic of soft lubrication is generated. We calculate this capillary-lubrication force, which is split into velocity-dependant and acceleration-dependant contributions. Furthermore, we analyze the variations induced by modifying the viscosity ratio between the two fluid layers, their thickness ratio, and the Bond number. Unlike standard elastic cases, where a repelling soft-lubrication lift force has been abundantly reported, the current fluid bilayer setting can also exhibit an attractive force due to the non-monotonic deflection of the fluid interface when varying the sublayer thickness. Besides, at high Bond numbers, the system's response becomes analogous to the one of a Winkler-like substrate with a viscous flow inside.
\end{abstract}
\maketitle

\section{\label{sec:intro}Introduction}
The movement of solid objects in viscous fluids has been the subject of detailed research in fluid sciences for more than a century~\cite{lamb1924hydrodynamics,Batchelor1967}. As opposed to particle motion in a bulk fluid~\cite{jeffery1915steady,collins1955steady,dean1963slow,o1964slow}, when an object moves in close proximity to a boundary, the resulting pressure field and the force exerted on the object are modified~\cite{o1967slow,goldman1967slow,cooley1969slow,jeffrey1981slow}. Such lubricated contacts have implications spanning over different domains, from tribology~\cite{hamrock2004fundamentals} to biomechanics of synovial fluids in joints~\cite{hou1992analysis,hlavavcek1993role} or the transport of cells in the blood~\cite{abkarian2002tank}. The understanding of single particle dynamics then further helps explaining the properties of clusters and suspensions~\cite{happel1983low,batchelor1970stress,batchelor1971stress,batchelor1976brownian,batchelor1977effect}. 

Over the past decades, dedicated research has explored the influence of soft boundaries on the motion of the particles to understand the role of boundary elasticity on hydrodynamic flow. The force and torque felt by the particle approaching the boundary have been calculated~\cite{leroy2011hydrodynamic,leroy2012hydrodynamic} and have been used to design the contactless rheological probes employed for soft materials~\cite{garcia2016micro,Basoli2018}. Surprisingly, as opposed to rigid boundaries, particles translating parallel to soft boundaries feel a repulsive lift force that arises out of the symmetry breaking induced by the elasticity of the wall~\cite{sekimoto1993mechanism,Beaucourt2004,skotheim2005soft,Weekley2006,urzay2007elastohydrodynamic,Snoeijer2013,salez2015elastohydrodynamics,Bouchet2015,Saintyves2016,Davies2018,Rallabandi2018,Vialar2019,zhang2020direct,Essink2021,bertin2022soft,Bureau2023,rallabandi2024fluid}. The scope has been further expanded to explore the influence of fluid compressibility~\cite{Balmforth2010}, fluid inertia~\cite{clarke2011elastohydrodynamics}, viscoelasticity of the boundary~\cite{Rallabandi2018}, and the inhomogeneities in slippage at the boundary~\cite{rinehart2020lift}. 

As the solids become softer, capillary stresses dominate over the material's bulk elasticity, and inner flows become increasingly important. The latter start to modify the force and torque generated. In the limit of point forces, previous research~\cite{aderogba1978action,nezamipour2021flow} has highlighted the pumping flow that can be observed when the interface deflection is accounted for. On the other hand, Leal and coworkers calculated the force felt by a  finite-sized sphere moving near a fluid interface, by utilizing Lorentz's Reciprocal Theorem, for the regime of a large gap as compared to the size of the sphere~\cite{lee1979motion,lee1980motion,lee1982motion,geller1986creeping}. Further developments included advancements in slender-body theory~\cite{yang1983particle} to explain the swimming of microorganisms near fluid interfaces~\cite{trouilloud2008soft,lopez2014dynamics}, as well as the formation of floating biofilms~\cite{desai2020biofilms}. A recent study on viscoelastic fluid substrates~\cite{hu2023effect} also highlighted that capillary interfaces could result in an attractive force instead of a repulsive lift one. 

While previous research has shown the importance and applicability of understanding the motion near a fluid interface, the characterization across different viscosity ratios and arbitrary layer thicknesses, for immiscible fluids, remains to be done. In the present article, we study the system of two rotating cylinders in close proximity to each other, separated by two viscous fluid layers. We calculate the force generated on one of the cylinders in the limit of small deformation of the fluid interface, as characterized by the capillary compliance. The article is organized as follows. We start by describing the viscocapillary lubrication problem at stake, followed by the theoretical methodology to obtain the different fields using perturbation analysis at small capillary compliance. We then discuss the implications of the deformable interface and the sub-layer flows on the force generated on the cylinder. 

\section{\label{sec:capillary_lubrication_theory}Capillary-lubrication theory}
\begin{figure}[h]
\begin{center}
\includegraphics[width=8cm]{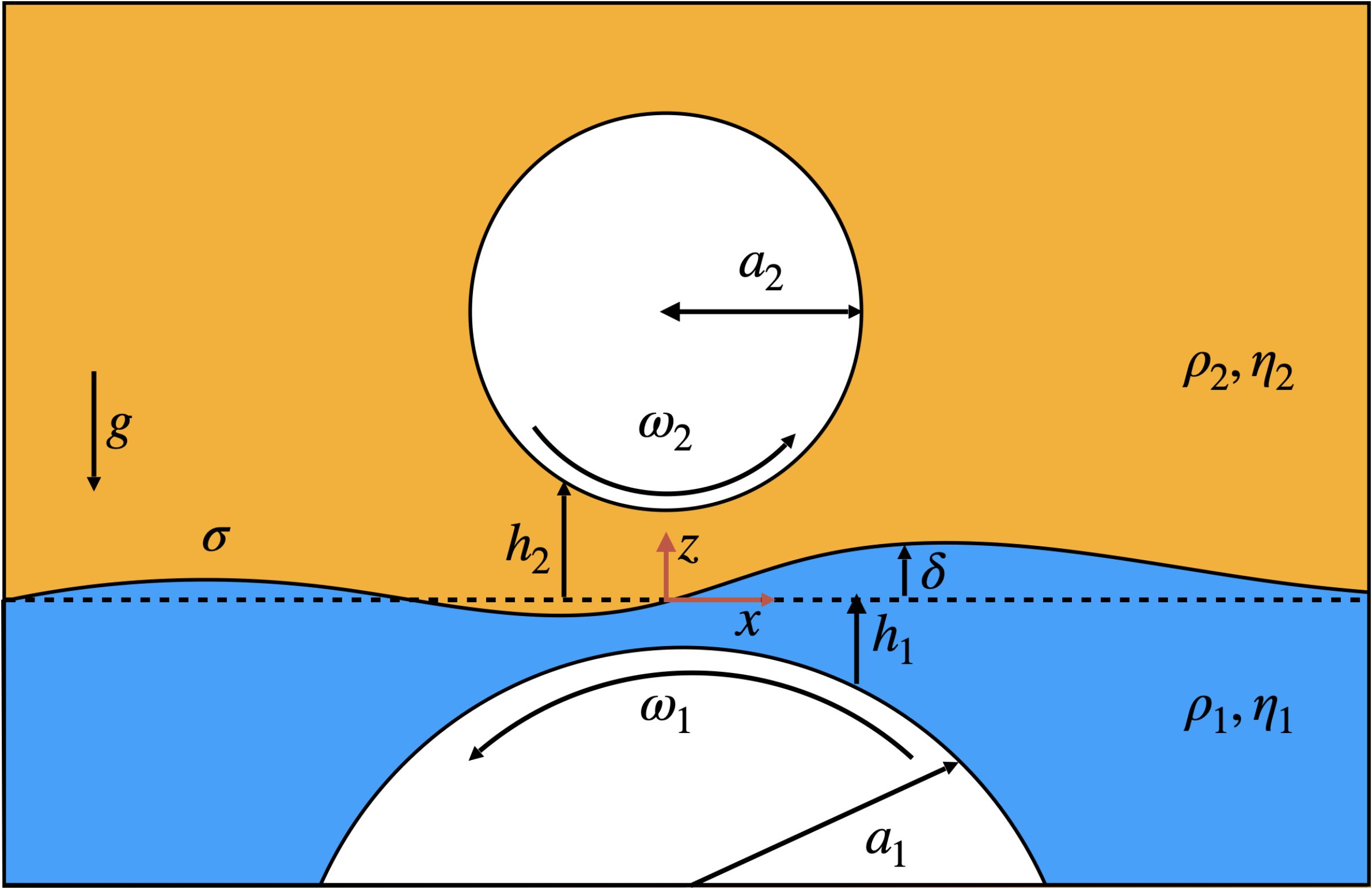}
\caption{Schematic of the system. Two rigid infinite cylinders rotate with prescribed velocities near a capillary interface between two incompressible Newtonian viscous fluids. The origin of spatial coordinates is located at the undeformed fluid interface ($z=0$) in the line joining the centers of mass of the cylinders ($x=0$). The deformed fluid interface is located at $z=\delta(x,t)$.}
\label{fig:1}
\end{center}
\end{figure}

We consider two rigid infinite cylinders of radii $a_1$ and $a_2$ rotating with prescribed time-dependent angular velocities $\omega_1$ and $\omega_2$ near a fluid interface, as shown in Fig.~\ref{fig:1}. The interface is characterized by its surface tension $\sigma$, and separates two incompressible Newtonian viscous fluids, with dynamic shear viscosities $\eta_1$ and $\eta_2$, as well as mass densities $\rho_1$ and $\rho_2$ (with $\rho_2<\rho_1$). The acceleration of gravity is denoted $g$. The thickness profiles $z=-h_1(x)$ and $z=h_2(x)$ of the bottom and top cylinders, depend on the horizontal position $x$. We denote by $z$ the vertical position and by $t$ the time.

\subsection{Governing equations}
We neglect fluid inertia and assume the typical thicknesses $d_i$, of the two fluids indexed by $i=1,2$, to be much smaller than the relevant horizontal length scales, defined by the hydrodynamic radii $\sqrt{2a_id_i}$~\cite{leroy2011hydrodynamic}, allowing us to invoke lubrication theory~\cite{Reynolds1886,Oron1997}. Introducing the excess pressure fields $p_i(x,z,t)$ with respect to the hydrostatic contributions, and the horizontal velocity fields $u_i(x,z,t)$, the incompressible Stokes equations thus read at leading lubrication order:
\begin{align}
\frac{\partial p_i}{\partial z} &= 0\ ,\label{eq:stokes1}\\
\frac{\partial p_i}{\partial x} &= \eta_i\frac{\partial^2 u_i}{\partial z^2}\ .\label{eq:stokes2}
\end{align}
In the near-contact region, in the limit of small gap, the shapes of the cylinders can be approximated by their parabolic expansions, as:
\begin{align} 
h_1(x) \simeq d_1+\frac{x^{2}}{2 a_1}~,\label{eq:h1} 
\end{align}
and:
\begin{align}  
h_2(x) \simeq d_2+\frac{x^{2}}{2 a_2}~.\label{eq:h2}
\end{align}
Finally, we close the equations by setting the flow boundary conditions. We impose no slip at the three interfaces alongside the balance of tangential and normal stresses at the fluid interface located at $z=\delta(x,t)$. Hence, at $z = -h_1$, one has:
\begin{align}
u_{1} = -\omega_1 a_1~, \label{eq:u1_bndcn}
\end{align}
at $z = h_2$, one has: 
\begin{align}
u_{2} = \omega_2 a_2~, \label{eq:u2_bndcn}
\end{align}
and at $z = \delta$, one has:  
\begin{align}
u_{2} & = u_1,\label{eq:vel_bndcn}\\
\eta_2\frac{\partial u_2}{\partial z} & = \eta_1\frac{\partial u_1}{\partial z}~, \label{eq:tangential_stress_balance}\\
p_2-p_{1}& \simeq\sigma \frac{\partial^2 \delta}{\partial x^2}+g\delta(\rho_2-\rho_1)~.\label{eq:laplace_dimensional}
\end{align}

Let us now non-dimensionalize the equations through: 
\begin{align*}
h_1(x)& = d_2H_1(X)~,   & h_2(x)& = d_2H_2(X)~,     & x & =lX~,  &  z & =d_2Z~, \\
t& = \frac{l}{c}T~, & u_1(x,z,t)& =cU_1(X,Z,T)~, &u_2(x,z,t)& =cU_2(X,Z,T)~,\\
p_1(x,t) &=\frac{\eta_2c l}{d_2^{\,2}}P_1(X,T)~, & p_2(x,t)&=\frac{\eta_2cl}{d_2^{\,2}}P_2(X,T)~, & \delta(x,t) & = d_2\Delta(X,T)~,
\end{align*}
with the upper hydrodynamic radius $l=\sqrt{2a_2d_2}$, and where $c$ represents a characteristic horizontal velocity scale, \textit{e.g.} $a_2\omega_2$. Moreover, the viscosity ratio is denoted by $M = \eta_1/\eta_2$. Using these dimensionless variables, Eqs.~(\ref{eq:h1},\ref{eq:h2}) become:
\begin{align} 
H_1(X)=\alpha+\frac{X^{2}}{\beta}~, \label{eq:H1} \\
H_2(X)=1+X^{2}~,\label{eq:H2}
\end{align}
where $\alpha = d_1/d_2$ and $\beta = a_1/a_2$ are the geometrical aspect ratios of the problem. The solutions of the dimensionless versions of Eqs.~(\ref{eq:stokes1},\ref{eq:stokes2}) are of the form: 
\begin{align} 
U_{1} &=\frac{P_{1}^{\prime} Z^{2}}{2 M}+C_{1} Z+C_{2}~, \label{eq:genU1}\\ 
U_{2} &=\frac{P_{2}^{\prime}}{2} Z^{2}+C_{3} Z+C_{4}~, \label{eq:genU1}
\end{align}
where the prime symbol corresponds to the partial derivative with respect to $X$, and where the coefficients $C_j$, with $j=1,2,3,4$, can be calculated by using the boundary conditions of Eqs.~(\ref{eq:u1_bndcn},\ref{eq:u2_bndcn},\ref{eq:vel_bndcn},\ref{eq:tangential_stress_balance}). Doing so, we obtain the velocity profiles: 
\begin{align}
\begin{split} \label{eq:genU1}
U_1 = {}& P_{1}^{\prime} \left\{\frac{Z^2-H_{1}^{2}}{2 M}+\frac{Z+H_{1}}{M(\Delta-H_{2})- (\Delta+H_{1})}\left[\frac{\Delta^{2}-H_{1}^{2}}{2 M}-\Delta\left(\Delta-H_{2}\right)\right]\right\}\\
&+P_{2}^{\prime} \left\{  \frac{(Z+H_1)(\Delta-H_2)^2}{2\left.\left[M\left(\Delta-H_{2}\right)- (\Delta+H_{1}\right)\right]}\right\}-\frac{(V_1+V_2)(Z+H_1)}{M(\Delta-H_{2})- (\Delta+H_{1})}-V_1~,
\end{split}\\
\begin{split} \label{eq:genU2}
U_2 = {}& -P_{1}^{\prime} \left\{\frac{(Z-H_2)(\Delta+H_1)^2}{2(M(\Delta-H_2)-(\Delta+H_1))}\right\}+P_{2}^{\prime} \left\{\frac{Z^2-H_2^2}{2}+(Z-H_2)\left[-\Delta+\frac{M(\Delta-H_2)^2}{2[M(\Delta-H_2)-(\Delta+H_1)]}\right]\right\}\\
&-\frac{M(V_1+V_2)(Z-H_2)}{M(\Delta-H_{2})- (\Delta+H_{1})}+V_2~\ ,
\end{split}
\end{align}
where $V_{i}=a_i\omega_i/c$.
We then calculate the flow rates within the two fluid films, as:  
\begin{multline} 
Q_{1}= \int_{-H_{1}}^{\Delta} U_{1} \textrm{d} Z =-P_{1}^{\prime} \frac{\left(\Delta+H_{1}\right)^{3}}{12M}\frac{4 M\left(\Delta-H_{2}\right)-\left(\Delta+H_{1}\right)}{M\left(\Delta-H_{2}\right)-\left(\Delta+H_{1}\right)}+\frac{P_{2}^{\prime}\left(H_{1}+\Delta\right)^{2}\left(H_{2}-\Delta\right)^{2}}{4\left[M\left(\Delta-H_{2}\right)-\left(\Delta+H_{1}\right)\right]} \\ 
 -V_1\left(\Delta+H_{1}\right)-\frac{\left(\Delta+H_{1}\right)^{2}(V_1+V_2)}{2\left[M\left(\Delta-H_{2}\right)-\left(\Delta+H_{1}\right)\right]}~, \label{eq:fluxQ1}
\end{multline}
\begin{multline} 
Q_{2}= \int_{\Delta}^{H_{2}} U_{2} \textrm{d} Z =\frac{P_{1}^{\prime}}{4}\frac{\left(\Delta-H_{2}\right)^{2}\left(\Delta+H_{1}\right)^{2}}{\left[M\left(\Delta-H_{2}\right)-\left(\Delta+H_{1}\right)\right]}-P_{2}^{\prime}\frac{\left(H_{2}-\Delta\right)^{3}}{12} \frac{M\left(\Delta-H_{2}\right)-4\left(\Delta+H_{1}\right)}{M\left(\Delta-H_{2}\right)-\left(\Delta+H_{1}\right)} \\ 
+V_2\left(H_{2}-\Delta\right)+\frac{M\left(\Delta-H_{2}\right)^{2}(V_1+V_2)}{2\left[M\left(\Delta-H_{2}\right)-\left(\Delta+H_{1}\right)\right]}\ , \label{eq:fluxQ2}
\end{multline}
which, by introducing ad-hoc auxiliary functions $F_i$ and $I_i$, can be rewritten in a more compact fashion, as:
\begin{align}
Q_1 = F_1(H_1,H_2,\Delta)P_1^{\prime}+F_2(H_1,H_2,\Delta)P_2^{\prime}+I_1(H_1,H_2,\Delta)~,\\
Q_2 = F_3(H_1,H_2,\Delta)P_1^{\prime}+F_4(H_1,H_2,\Delta)P_2^{\prime}+I_2(H_1,H_2,\Delta)~.
\end{align}

The thin-film equations for this system can be derived by invoking volume conservation in the two fluid layers, as: 
\begin{align} 
\frac{\partial \Delta}{\partial T}+Q_1^\prime = 0~,\label{eq:thinfilm1}\\ 
\frac{\partial \Delta}{\partial T}-Q_2^\prime = 0~. \label{eq:thinfilm2}
\end{align}
Finally, Eq.~(\ref{eq:laplace_dimensional}) reads in dimensionless form:
\begin{align}
\Delta^{\prime\prime}-\textrm{Bo}\Delta=\kappa\left(P_2-P_1\right)~, \label{eq:laplaceentire}
\end{align}
where $\textrm{Bo} = (l/l_{\textrm{c}})^2$ denotes the Bond number of the problem which compares the relevant dynamical horizontal length scale $l$ to the capillary length $l_{\textrm{c}}=\sqrt{\sigma/[g(\rho_1-\rho_2)])}$. The dimensionless compliance of the fluid interface is denoted by $\kappa= \textrm{Ca}/\epsilon^3$, where $\textrm{Ca} = \eta_2c/\sigma$ is a capillary number and $\epsilon=d_2/l$ is a small lubrication parameter. The problem has three unknown fields: $\Delta$, $P_1$ and $P_2$. These obey the set of three coupled differential equations given by Eqs.~(\ref{eq:thinfilm1}-\ref{eq:laplaceentire}), together with the following spatial boundary conditions: $P_{i} \rightarrow 0$ and $\Delta \rightarrow 0$ at $X\rightarrow \pm\infty$.

\section{Perturbation analysis}
Following the approach of previous soft-lubrication studies~\cite{sekimoto1993mechanism,skotheim2005soft, urzay2007elastohydrodynamic, salez2015elastohydrodynamics,pandey2016lubrication}, we assume that $\kappa\ll1$ and perform an expansion of the fields up to first order in $\kappa$, as: 
\begin{align}
\Delta &\simeq0+ \kappa \Delta_1~, \label{eq:21}\\
P_1 &\simeq P_{10}+\kappa P_{11}~, \label{eq:22}\\
P_2 &\simeq P_{20}+\kappa P_{21}~,\\
U_1 &\simeq U_{10}+\kappa U_{11}~,\\
U_2 &\simeq U_{20}+\kappa U_{21}~, 
\label{eq:23}
\end{align}
where $\kappa\Delta_1$ is the deformation profile of the fluid interface at first order in $\kappa$, $\kappa^{j}P_{ij}$ the excess pressure field in layer $i$ at perturbation order $j$, and $\kappa^{j}U_{ij}$ the velocity field in layer $i$ at perturbation order $j$. Given the respective symmetries of the fields at each order in $\kappa$, it is more convenient to focus only on the $X>0$ domain, and impose the following equivalent spatial boundary conditions: $P_{i0} = 0$, $P_{i1}^{\prime} = 0$ and $\Delta_1=0$ at $X = 0$, as well as $P_{ij} \rightarrow 0$ and $\Delta_1 \rightarrow 0$ at $X\rightarrow+ \infty$.

\subsection{Zeroth-order solution}
At zeroth order in $\kappa$, the fluid interface is undeformed. Equations~(\ref{eq:thinfilm1},\ref{eq:thinfilm2}) then lead to the following coupled ordinary differential equations for the two pressure fields:  
\begin{align} 
F_{10}P_{10}^{\prime} +F_{20}P_{20}^{\prime} =k_{1}-I_{10}~, \\ 
F_{30}P_{10}^{\prime} +F_{40}P_{20}^{\prime} =k_{2}-I_{20}~, 
\end{align}
where the $k_i$ are integration constants, and where we have evaluated the above auxiliary functions at zeroth order in $\kappa$, as: 
\begin{align}
\begin{split}
F_{10} =-\frac{H_{1}^{3}}{12 M}\frac{4 M H_{2}+H_{1}}{M H_{2}+H_{1}}~,
\end{split}
\begin{split}
F_{20} = F_{30} = -\frac{H_{1}^{2} H_{2}^{2}}{4\left(M H_{2}+H_{1}\right)}~, \qquad
\end{split}
\begin{split}
F_{40} = -\frac{H_{2}^{3}}{12}\frac{M H_{2}+4 H_{1}}{M H_{2}+H_{1}}~,
\end{split}
\end{align}
\begin{align}
\begin{split}
I_{10}=\frac{H_{1}^{2}(V_1+V_2)}{2\left(M H_{2}+H_{1}\right)}-V_1 H_{1}~, \hspace{-25pt}
\end{split} 
\begin{split}
\hspace{-50pt}I_{20}=V_2 H_{2}-\frac{M H_{2}^{2}(V_1+V_2)}{2\left(M H_{2}+H_{1}\right)}~.
\end{split}
\end{align}
The derivatives of the pressure fields can then be evaluated from the above expressions, to give: 
\begin{align} 
& P_{10}^{\prime}=\frac{F_{40}\left(k_{1}-I_{10}\right)-F_{20}\left(k_{2}-I_{20}\right)}{F_{10} F_{40}-F_{20} F_{30}}~, \label{eq:P10_grad} \\
& P_{20}^{\prime}=\frac{F_{10}\left(k_{2}-I_{20}\right)-F_{30}\left(k_{1}-I_{10}\right)}{F_{10} F_{40}-F_{20} F_{30}}~. \label{eq:P20_grad}
\end{align}
The latter equations can be integrated, \textit{e.g.} with Mathematica or an explicit finite-difference numerical method -- both giving identical results. The obtained solutions will be analyzed in the discussion section.

\subsection{First-order solution}
According to Eq.~(\ref{eq:laplaceentire}), the zeroth-order pressure fields calculated above lead to a deflection of the interface at first order in $\kappa$, which satisfies:
\begin{align}
\Delta_1^{\prime\prime}-\textrm{Bo}\Delta_1 = P_{20}-P_{10} ~.\label{eq:Laplace_nondim}
\end{align}
The formal solution of Eq.~\eqref{eq:Laplace_nondim} satisfying the above boundary conditions reads:
\begin{align}
\Delta_1 (X,T)=  \Delta_{\textrm{c}}(T)\sinh\left(X\sqrt{\textrm{Bo}}\right)-\frac{1}{\sqrt{\textrm{Bo}}}\int_0^X\textrm{d}Y\,\left[P_{20}(Y,T)-P_{10}(Y,T)\right]\sinh\left[(Y-X)\sqrt{\textrm{Bo}}\right]~,
\label{gensol}
\end{align}
where $\Delta_{\textrm{c}}(T)=\left(1/\sqrt{\textrm{Bo}}\right)\int_0^{\infty}\textrm{d}Y\,\left[P_{10}(Y,T)-P_{20}(Y,T)\right]\exp\left(-Y\sqrt{\textrm{Bo}}\right)$. This solution can be numerically evaluated for fixed parameters Bo, $M$, $\alpha$ and $\beta$.

Then, from the obtained deflection at first order in $\kappa$, one can calculate the pressure fields at first order in $\kappa$, as explained hereafter. To begin with, the auxiliary functions are evaluated at first order in $\kappa$, as: 
\begin{align}
F_{n}&\simeq F_{n0} + \kappa \Delta_1\frac{\partial F_n}{\partial \Delta}\bigg\rvert_{\Delta = 0}~,\\
I_{m}&\simeq I_{m0} + \kappa \Delta_1\frac{\partial I_m}{\partial \Delta}\bigg\rvert_{\Delta = 0}\ ,
\end{align}
for $n=1,2,3,4$ and $m=1,2$. Introducing $G_{n0} = \frac{\partial F_n}{\partial \Delta}\rvert_{\Delta = 0}$, and $E_{m0} = \frac{\partial I_m}{\partial \Delta}\rvert_{\Delta = 0}$, and expending the fluxes as $Q_i\simeq Q_{i0}+\kappa Q_{i1}$, one gets the first-order corrections to the fluxes: 
\begin{align} 
& Q_{11} \simeq F_{10} P_{11}^{\prime}+\Delta_1 G_{10}P_{10}^{\prime}+F_{20}P_{21}^{\prime}+\Delta_1 G_{20}P_{20}^{\prime}+\Delta_1 E_{10}~, \\ 
& Q_{21} \simeq F_{30}P_{11}^{\prime}+\Delta_1 G_{30} P_{10}^{\prime}+F_{40} P_{21}^{\prime}+\Delta_1 G_{40} P_{20}^{\prime}+\Delta_1 E_{20}~.
\end{align}
Furthermore, the thin-film equations and the $X=0$ boundary conditions imply: 
\begin{align} 
Q_{11}&=-\int_{0}^{X} \frac{\partial \Delta_1}{\partial T} \textrm{d} X~, \\ 
Q_{21}&=\int_{0}^{X} \frac{\partial \Delta_1}{\partial T} \textrm{d} X~.
\end{align}
Combining the last four equations leads to: 
\begin{align} 
& F_{10}P_{11}^{\prime}+F_{20} P_{21}^{\prime}=-\mathcal{K}-\mathcal{J}_1~, \label{eq:firstorder_1}\\ 
& F_{30} P_{11}^{\prime}+F_{40}P_{21}^{\prime}=\mathcal{K}-\mathcal{J}_2~, \label{eq:firstorder_2}
\end{align}
with:
\begin{align} 
\mathcal{J}_1 &= \Delta_1 \left(G_{10} P_{10}^{\prime}+ G_{20} P_{20}^{\prime}+ E_{10}\right)~,\\
\mathcal{J}_2 &= \Delta_1 \left( G_{30} P_{10}^{\prime}+ G_{40} P_{20}^{\prime}+ E_{20}\right)~, \\
\mathcal{K} &=\int_0^X\frac{\partial \Delta_1}{\partial T}\textrm{d}X ~.
\end{align}
Decoupling the equations, leads to:
\begin{align}
P_{11}^{\prime}&=\frac{F_{40}\mathcal{H}_1-F_{20}\mathcal{H}_2}{F_{10}F_{40}-F_{20}F_{30}}~, \label{eq:p_11grad}\\
P_{21}^{\prime}&=\frac{F_{10}\mathcal{H}_2-F_{30}\mathcal{H}_1}{F_{10}F_{40}-F_{20}F_{30}}~, \label{eq:p_21grad}
\end{align}
with $\mathcal{H}_1=-\mathcal{K}-\mathcal{J}_1$ and $\mathcal{H}_2=\mathcal{K}-\mathcal{J}_2$. These can be numerically integrated over $X$, for fixed parameters Bo, $M$, $\alpha$ and $\beta$, using the far-field boundary conditions. The obtained solutions will be analyzed in the discussion section.

We conclude this section by an important remark. In the above expressions, we see that $\mathcal{K}$ involves the accelerations of the cylinders, while $\mathcal{J}_1$ and $\mathcal{J}_2$ both involve squared velocities instead. Moreover, the first-order corrections of the pressure fields are linear combinations of $\mathcal{K}$ and $\mathcal{J}_i$, which is reminiscent of past soft-lubrication studies~\cite{kaveh2014hydrodynamic,bertin2022soft,jha2023capillary}. Therefore, in order to address these two independent forcing modes later on, we split $\mathcal{H}_1$ and $\mathcal{H}_2$ into their: i) squared-velocity-dependant contributions $(\mathcal{H}_i)_{U^2}$, generically denoted by the subscript ``$U^2$", and referred to as ``lift" terms ; and ii) acceleration-dependant contributions $(\mathcal{H}_i)_{\dot{U}}$, generically denoted by the subscript ``$\dot{U}$", and referred to as ``inertial" terms.  These contributions read:  
\begin{align}
(\mathcal{H}_1)_{U^2} &= -\mathcal{J}_1~,\\
(\mathcal{H}_2)_{U^2} &= -\mathcal{J}_2~,\\
(\mathcal{H}_1)_{\dot{U}} &= -\mathcal{K}~,\\
(\mathcal{H}_2)_{\dot{U}} &= \mathcal{K}~. 
\end{align}

\section{Discussion}
Hereafter, keeping $\beta\gg 1$ in order to approach the situation of a cylinder moving near a thin, supported and flat fluid film, we discuss the zeroth-order and first-order solutions, and investigate the influence of the three other key dimensionless parameters of the problem: the viscosity ratio $M$, the gap ratio $\alpha$, and the Bond number Bo.

\subsection{Zeroth-order pressure}
\begin{figure}
\begin{subfigure}[b]{0.45\textwidth}
\caption{}
\centering
\includegraphics[width=8cm]{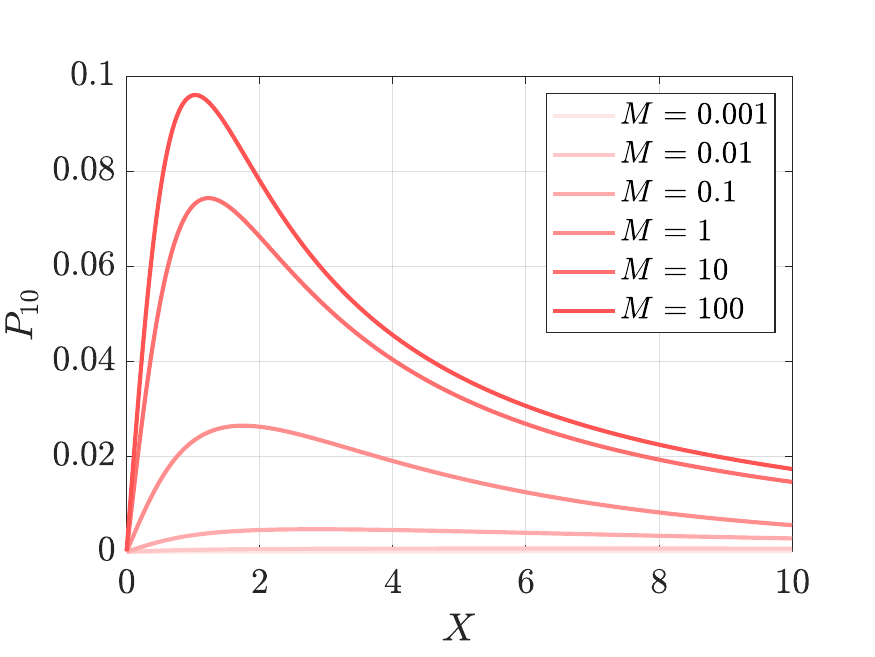}   
\end{subfigure}
\hfill
\begin{subfigure}[b]{0.45\textwidth}
\caption{}
\centering
\includegraphics[width=8cm]{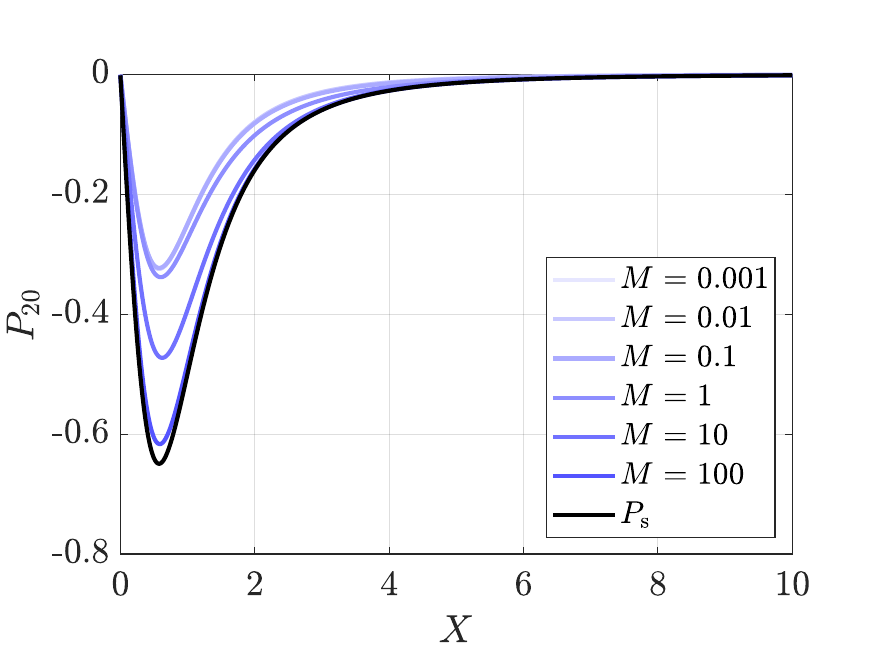}
\end{subfigure}
\caption{Zeroth-order excess pressure fields, $P_{10}$ (a) and $P_{20}$ (b), as functions of the horizontal coordinate $X$, as evaluated from Eqs.~(\ref{eq:P10_grad},\ref{eq:P20_grad}) for $\alpha = 15$ and $\beta = 99$, with $V_1 = 1$ and $V_2 = 1$, and different values of $M$ as indicated. The black line in panel (b) represents the pressure profile $P_{\textrm{s}}$ generated near a solid boundary~\cite{jeffrey1981slow}.}
\label{fig:p0_Mvariation}
\end{figure}
The zeroth-order excess pressure fields in the two layers are computed from Eqs.~(\ref{eq:P10_grad},\ref{eq:P20_grad}) and plotted in Fig.~\ref{fig:p0_Mvariation}, for $\alpha = 15$ and $\beta = 99$, and for different values of the viscosity ratio $M$. For comparison, we also show the pressure $P_{\textrm{s}}(X) = -2V_2X/(1+X^2)^2$~\cite{jeffrey1981slow} generated if the fluid interface was replaced by a no-slip solid boundary. The pressure fields in both layers appear to have opposite signs. Furthermore, due to the allowed flow in the bottom layer, the pressure generated in the top layer is lower than if the interface was a no-slip solid substrate. However, increasing $M$, \textit{i.e.} increasing the viscosity of the bottom layer with respect to the one in the top layer increases the pressure magnitude in both layers. Eventually, at large $M$, the pressure field in the top layer saturates towards $P_{\textrm{s}}$, as expected.

\begin{figure}
\begin{subfigure}[b]{0.45\textwidth}
\caption{}
\centering
\includegraphics[width=8cm]{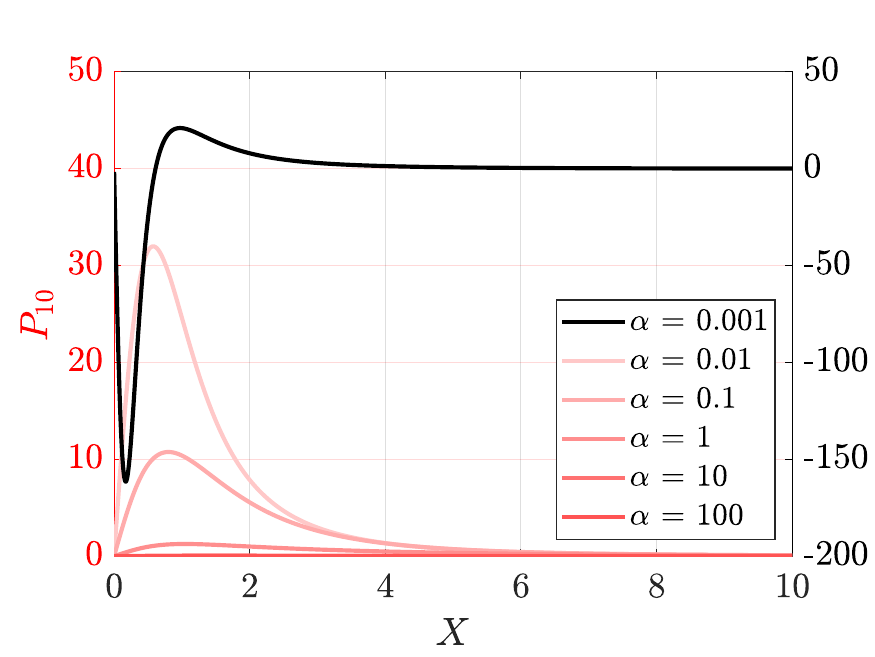}
\end{subfigure}
\hfill
\begin{subfigure}[b]{0.45\textwidth}
\centering
\includegraphics[width=8cm]{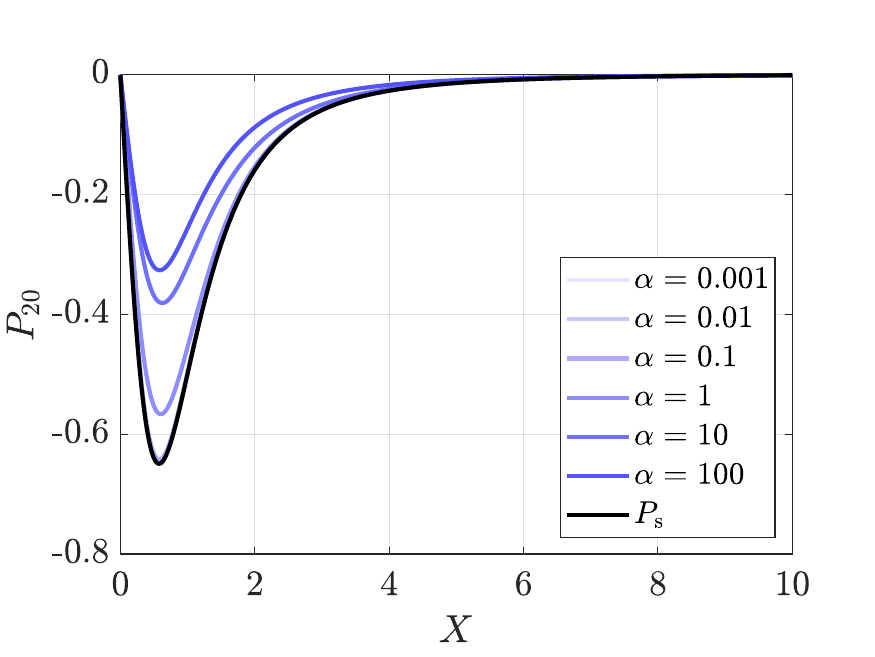}
\end{subfigure}
\caption{Zeroth-order excess pressure fields, $P_{10}$ (a) and $P_{20}$ (b), as functions of the horizontal coordinate $X$, as evaluated from Eqs.~(\ref{eq:P10_grad},\ref{eq:P20_grad}) for $M = 2$ and $\beta = 99$, with $V_1 = 1$ and $V_2 = 1$, and different values of $\alpha$ as indicated. The black line in panel (a) represents the top-layer pressure field for $\alpha = 0.001$, with the appropriate scale on the right $y$-axis. The black line in panel (b) represents the pressure profile $P_{\textrm{s}}$ generated near a solid boundary~\cite{jeffrey1981slow}.}
\label{fig:p0_alphavariation}
\end{figure}

In Fig.~\ref{fig:p0_alphavariation}, we investigate the effect of the ratio $\alpha$ between the bottom-layer and top-layer thicknesses. Reducing $\alpha$ increases the magnitude of the pressure generated in the top layer, as expected due to the reducing flow ability in the bottom layer. Once again, the curves eventually saturate towards the no-slip solid pressure $P_{\textrm{s}}$. Interestingly, at some point, the decrease of $\alpha$ leads to a sign change for the bottom-layer pressure. The transition point of such a sign change depends (not shown) upon the chosen values of $M$ and $\beta$.  

\subsection{Interface deflection}
\begin{figure}
\centering
\includegraphics[width=12cm]{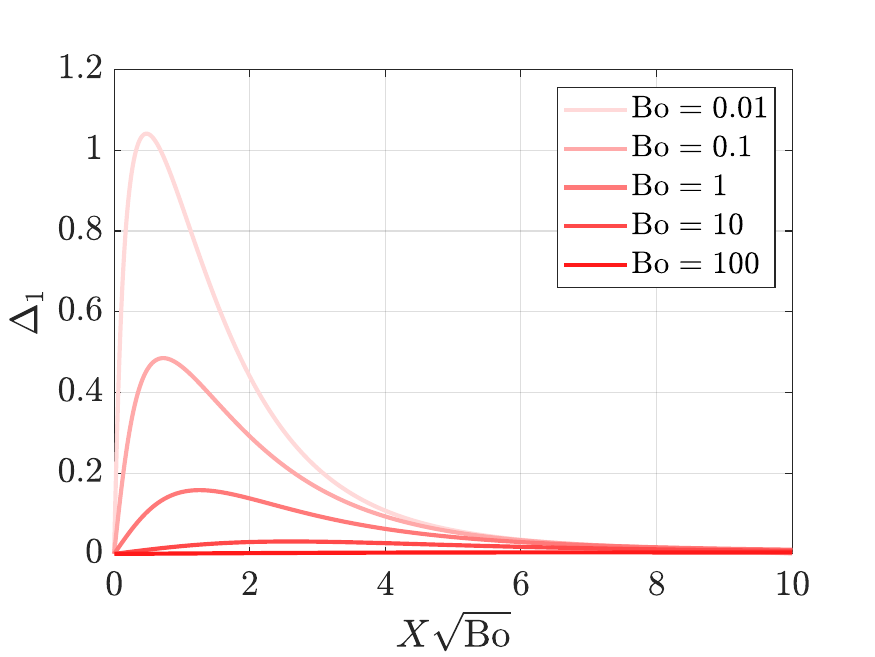}
\caption{First-order interface deflection profile $\Delta_1$ as a function of rescaled horizontal coordinate $X\sqrt{\textrm{Bo}}$, as calculated from Eq.~(\ref{gensol}) with $M = 2$, $\alpha = 15$, $\beta = 99$, $V_1 = 0$, $V_2 = 1$, and for several Bo values as indicated.}
\label{fig:interface_XBo}
\end{figure}

The first-order interface deflection field is calculated using Eq.~(\ref{gensol}) and plotted in Fig.~\ref{fig:interface_XBo} for several Bo values. While the horizontal range and the magnitude are both affected by Bo, the former can be absorbed into a rescaled horizontal coordinate $X\sqrt{\textrm{Bo}}$. This is characteristic of problems involving capillarity and a direct consequence of the Young-Laplace condition of Eq.~(\ref{eq:Laplace_nondim}). Besides, the deflection magnitude decreases as Bo is increased, since gravitational resistance towards interface deformation is increased. However, the decrease does not seem to follow a simple scaling law with Bo. 

\begin{figure}
\begin{subfigure}[b]{0.45\textwidth}
\caption{}
\centering
\includegraphics[width=8cm]{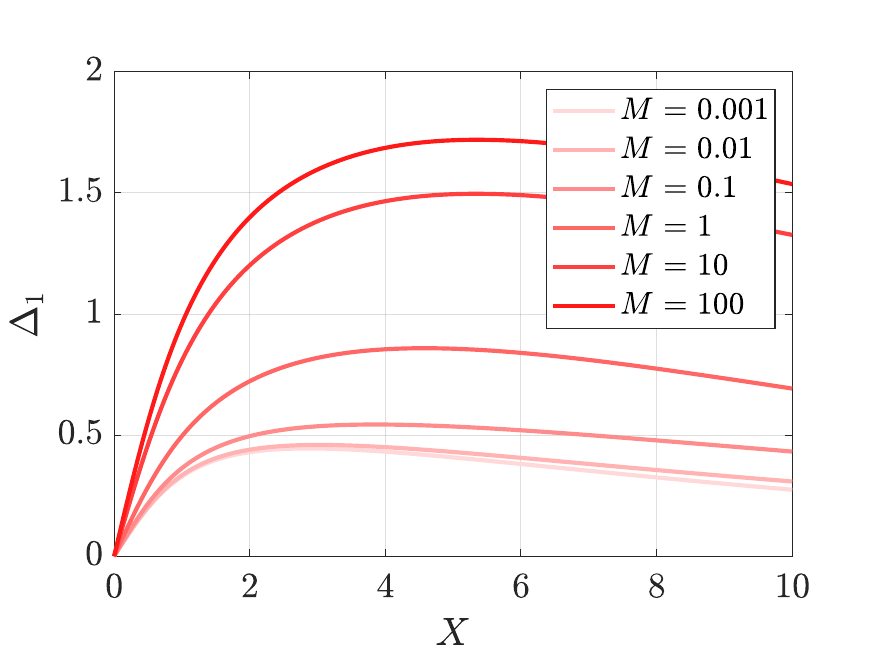}
\end{subfigure}
\hfill
\begin{subfigure}[b]{0.45\textwidth}
\caption{}
\centering
\includegraphics[width=8cm]{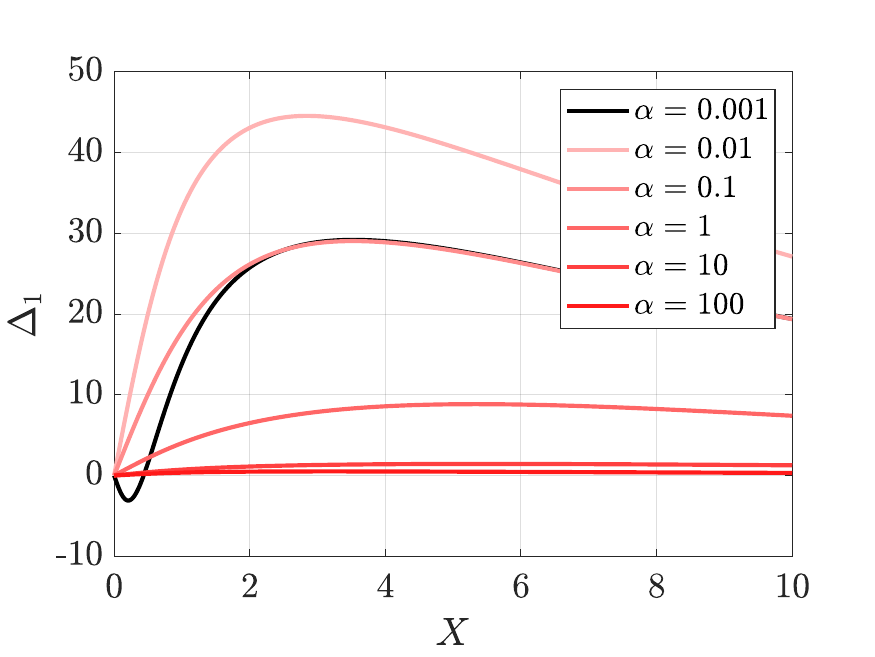}
\end{subfigure}
\caption{First-order interface deflection profile $\Delta_1$ as a function of horizontal coordinate $X$, as calculated from Eq.~(\ref{gensol}) with Bo~$= 0.01$, $\beta = 99$, $V_1= 0$ and $V_2 = 1$, for: (a) $\alpha = 15$ and varying $M$ as indicated; and (b) $M = 2$ and varying $\alpha$ as indicated.}
\label{fig:interface_variation}
\end{figure}

Let us now investigate the influence of the viscosity ratio $M$ and gap ratio $\alpha$. The results are plotted in Fig.~\ref{fig:interface_variation}. As $M$ increases, the zeroth-order pressure fields increase in magnitude monotonically, leading to a corresponding increase in the magnitude of the interface deflection. Similarly, decreasing $\alpha$ increases the magnitude of the interface deflection. However, the sign change for the bottom-layer pressure field observed previously at small $\alpha$ leads to an intricate behaviour of the interface deflection profile. Further decreasing (not shown) $\alpha$ can even lead to a complete sign flip of the interface deflection. 

\subsection{First-order pressure}
\begin{figure}[h]
\includegraphics[width=10cm]{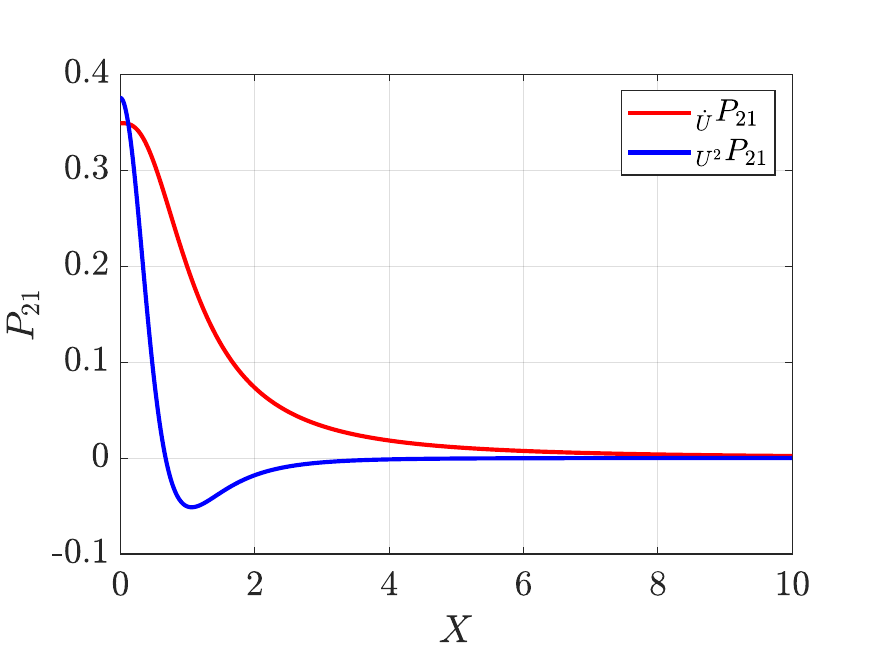}
\caption{First-order pressure correction $P_{21}$ in the top layer as a function of horizontal coordinate $X$, as obtained from numerically solving Eqs.~(\ref{eq:p_11grad},\ref{eq:p_21grad}) for  $M = 2$, $\alpha = 15$, $\beta = 99$, and $\textrm{Bo} = 0.01$. Both the lift term $_{U^2}P_{21}$ (blue) for $V_1 = 0$ and $V_2 = 1$, and inertial term $_{\dot{U}}P_{21}$ (red) for $\dot{V_1} = 0$ and $\dot{V_2} = 1$, are shown.}
\label{fig:p1} 
\end{figure}

Integrating Eqs.~(\ref{eq:p_11grad},\ref{eq:p_21grad}) allows us to find the first-order pressure corrections $P_{i1}=_{U^2}P_{i1}+\,_{\dot{U}}P_{i1}$, which are separated into two different contributions as mentioned beforehand: i) lift terms $_{U^2}P_{i1}$ ; and ii) inertial terms $_{\dot{U}}P_{i1}$. We further stress that our discussion below focuses only on the top layer, as our goal is to eventually calculate the force generated on the top cylinder. The lift and inertial terms in the top layer are shown in Fig.~\ref{fig:p1} for a given set of parameters. As we see, they typically push the cylinder away from the interface. Besides, as the Bond number Bo is increased, one observes (not shown) a decrease in both the lift and inertial terms, which is due to the reducing interface deflection observed above.

\begin{figure}
\begin{subfigure}[b]{0.45\textwidth}
\caption{}
\centering
\includegraphics[width=8cm]{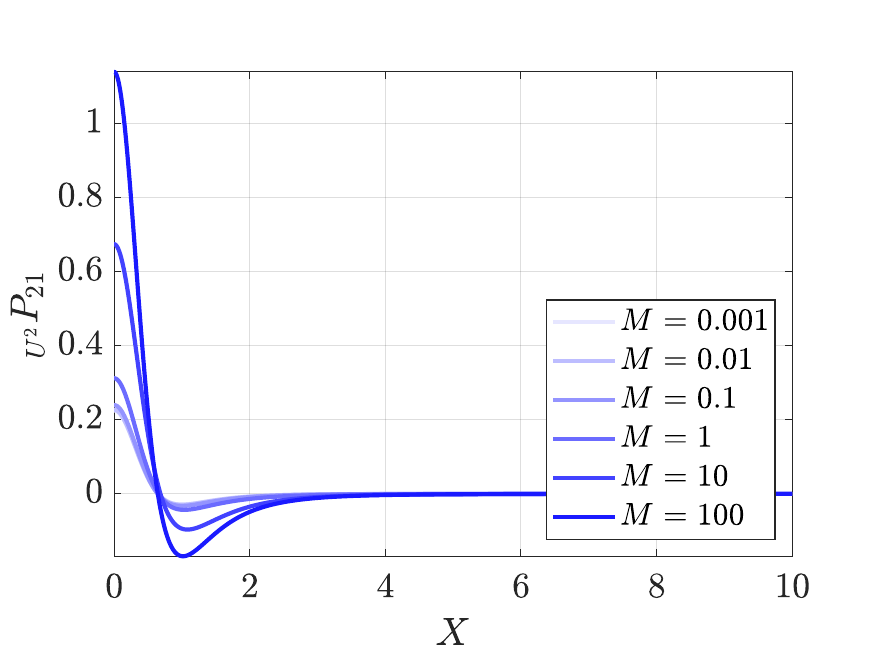}
\end{subfigure}
\hfill
\begin{subfigure}[b]{0.45\textwidth}
\caption{}
\centering
\includegraphics[width=8cm]{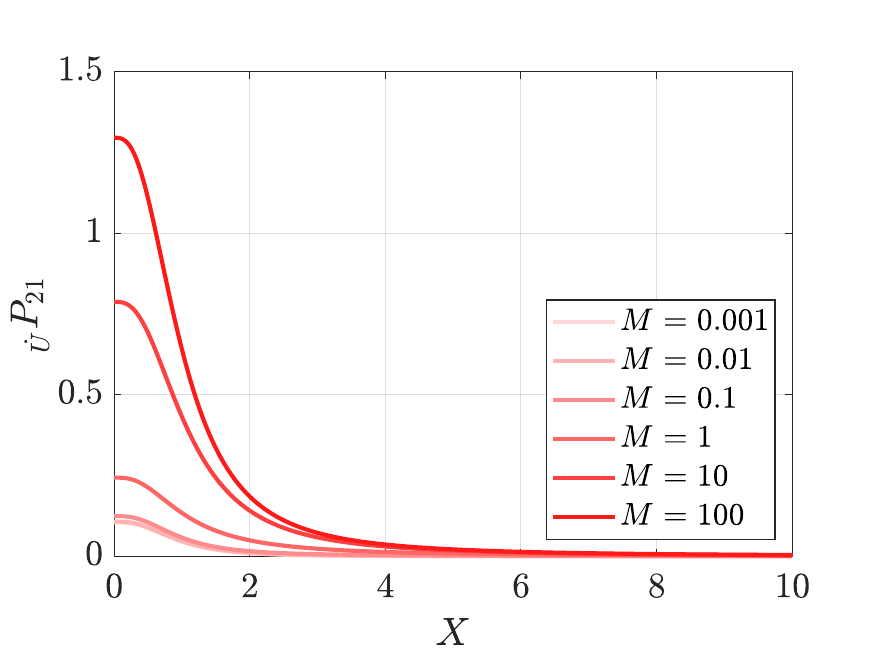}
\end{subfigure}
\caption{First-order pressure correction $P_{21}$ in the top layer as a function of horizontal coordinate $X$, as obtained from numerically solving Eqs.~(\ref{eq:p_11grad},\ref{eq:p_21grad}) with $\alpha = 15$, $\beta = 99$, and $\textrm{Bo} = 0.01$, and for various $M$ as indicated. Both the lift term $_{U^2}P_{21}$ (a) with $V_1 = 0$ and $V_2 = 1$, and inertial term $_{\dot{U}}P_{21}$ (b) with $\dot{V_1} = 0$ and $\dot{V_2} = 1$, are shown.}
\label{fig:p1_Mvariation}
\end{figure}

\begin{figure}
\begin{subfigure}[b]{0.45\textwidth}
\caption{}
\centering
\includegraphics[width=8cm]{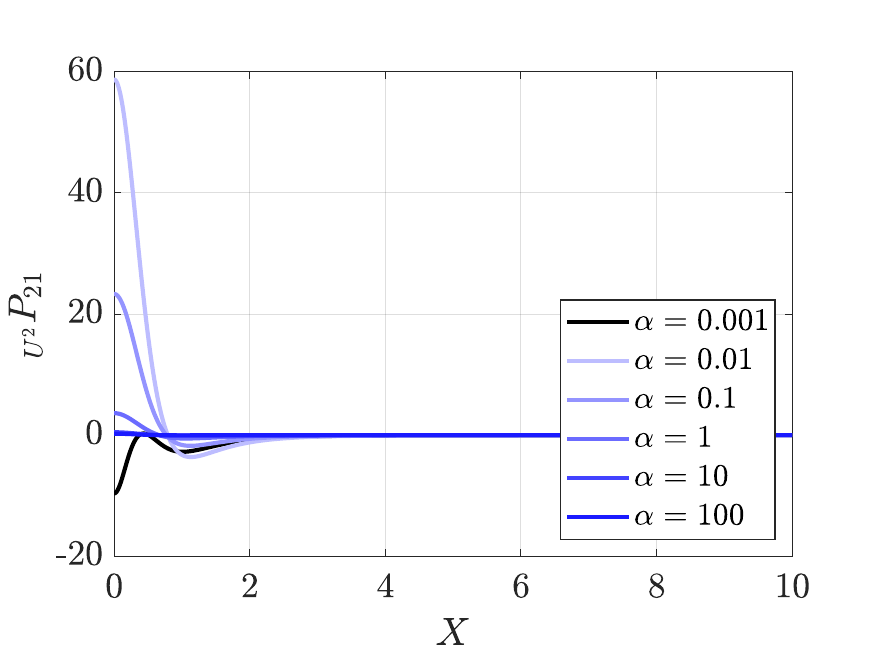}
\end{subfigure}
\hfill
\begin{subfigure}[b]{0.45\textwidth}
\caption{}
\centering
\includegraphics[width=8cm]{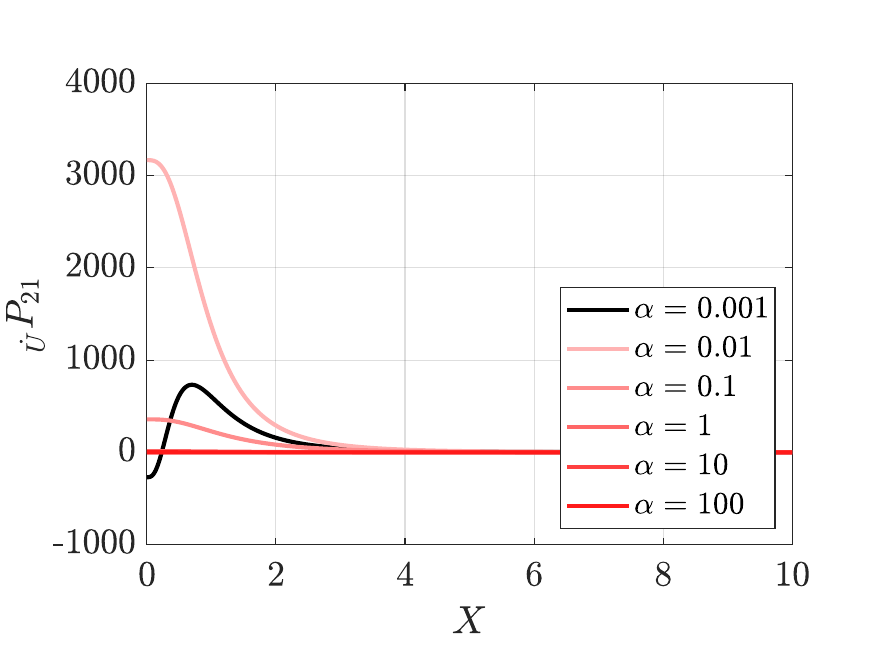}
\end{subfigure}
\caption{First-order pressure correction $P_{21}$ in the top layer as a function of horizontal coordinate $X$, as obtained from numerically solving Eqs.~(\ref{eq:p_11grad},\ref{eq:p_21grad}) with $M = 2$, $\beta = 99$, and $\textrm{Bo} = 0.01$, and for various $\alpha$ as indicated. Both the lift term $_{U^2}P_{21}$ (a) with $V_1 = 0$ and $V_2 = 1$, and inertial term $_{\dot{U}}P_{21}$ (b) with $\dot{V_1} = 0$ and $\dot{V_2} = 1$, are shown.}
\label{fig:p1_alpha_variation}
\end{figure}

Moreover, the effects of the viscosity ratio $M$ and gap ratio $\alpha$ are presented in Figs.~\ref{fig:p1_Mvariation} and~\ref{fig:p1_alpha_variation}. Increasing $M$, or decreasing $\alpha$, increases the magnitude of both the lift and inertial terms. This is explained once again by the reducing flow ability in the bottom layer. However, interestingly, for very small $\alpha$ values, where the zeroth-order pressure field in the bottom layer changes sign, the first-order pressure field in the top layer reduces in magnitude. 

\subsection{Capillary-lubrication force}
\begin{figure}
\begin{subfigure}[b]{0.45\textwidth}
\caption{}
\centering
\includegraphics[width=8cm]{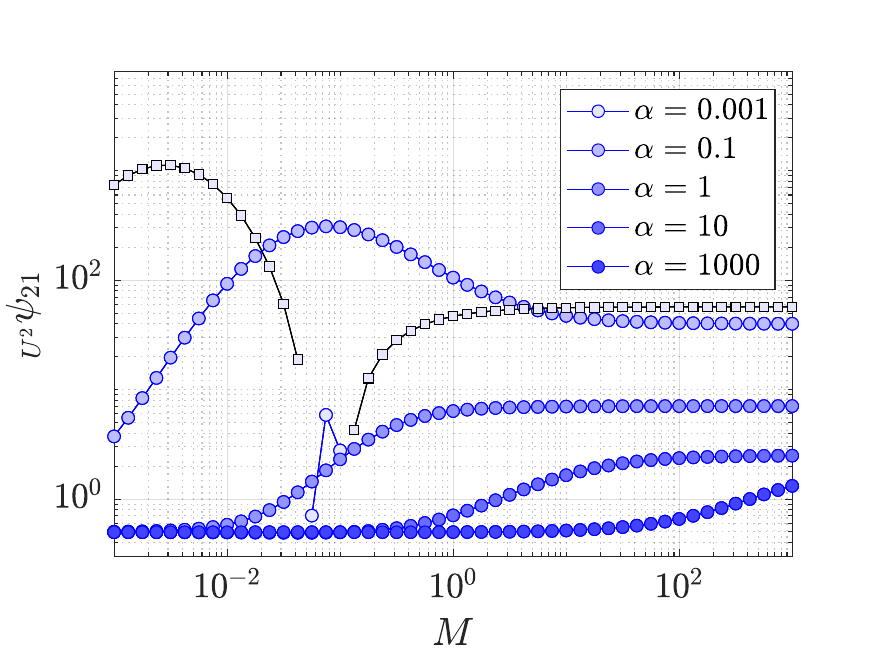}
\end{subfigure}
\hfill
\begin{subfigure}[b]{0.45\textwidth}
\caption{}
\centering
\includegraphics[width=8cm]{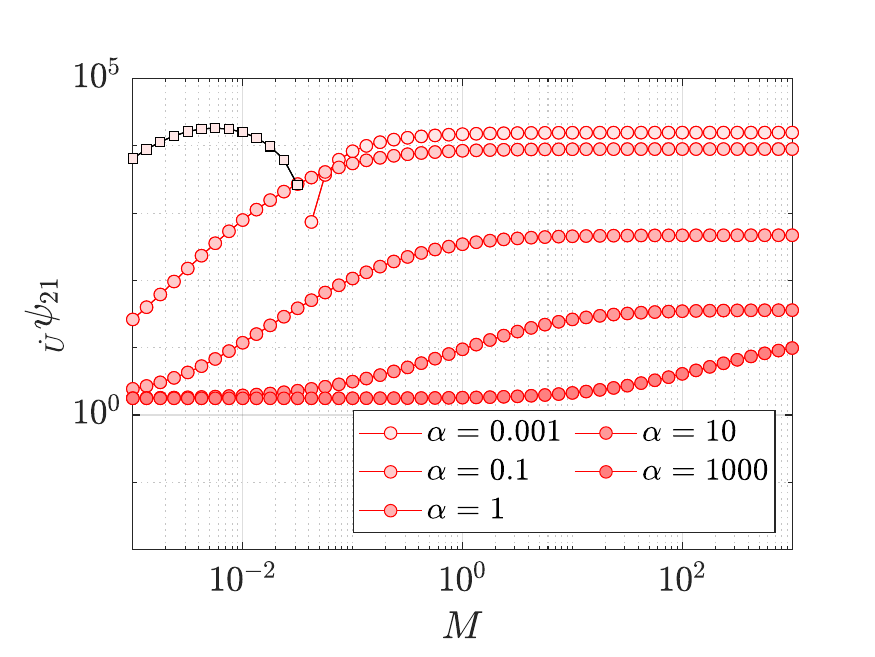}
\end{subfigure}
\caption{Lift coefficient $_{U^2}\psi_{21}$ (a) and inertial coefficient $_{\dot{U}}\psi_{21}$ (b) of the first-order normal force per unit length (see Eq.~(\ref{eq:normal_force})) exerted on the top cylinder as functions of the viscosity ratio $M$ and for various gap ratios $\alpha$. These coefficients were obtained from numerical integration of $P_{21}$ along $X$, as computed for Bo $= 0.01$, $\beta=99$, and for either $V_1 = 0$, $V_2 = 1$ (a), or $\dot{V_1} = 0$, $\dot{V_2} = 1$ (b). The squares denote the absolute values in the case of negative values.}
\label{fig:psi11_psi12}
\end{figure}
In dimensional units, the normal force per unit length $F$ felt by the top cylinder can be found by integrating the pressure field $p_2$ in the top layer along the horizontal coordinate $x$. At zeroth order, the force is null by symmetry of the excess pressure fields. At leading order in perturbation, one thus has:
\begin{align} 
 F = \int_{-\infty}^\infty p_2 \textrm{d}x \simeq\frac{\eta_2^2\omega_2^2a_2^2}{\sigma}\left(\frac{a_2}{d_2}\right)^{5/2}\, _{U^2}\psi_{21}+\frac{\eta_2^2\dot{\omega}_2a_2^2}{\sigma}\left(\frac{a_2}{d_2}\right)^{2}\, _{\dot{U}}\psi_{21}~,    \label{eq:normal_force}   
\end{align}
where $_{U^2}\psi_{21} = (2^{7/2}/V_2^{\,2})\int_{0}^{\infty}\, _{U^2}P_{21} \textrm{d}X $ and $ _{\dot{U}}\psi_{21} = (2^{4}/\dot{V_2})\int_{0}^{\infty} \,_{\dot{U}}P_{21}$ are dimensionless coefficients corresponding to the lift and inertial terms, respectively. These two coefficients are plotted in Fig.~\ref{fig:psi11_psi12} as functions of the viscosity ratio $M$ and for various gap ratios $\alpha$. Two main comments can be made on the results. First, the coefficients vary by orders of magnitude upon changing the two ratios, which indicates the possibility of tuning the conditions to modify, control and optimize the capillary-lubrication effects. Moreover, in direct contrast to classical elastic soft lubrication~\cite{Bureau2023}, the coefficients signs can be changed too, as already reported for the lift force in a recent study on viscoelastic fluid substrates~\cite{hu2023effect}.

\begin{figure}[h]
\includegraphics[width=12cm]{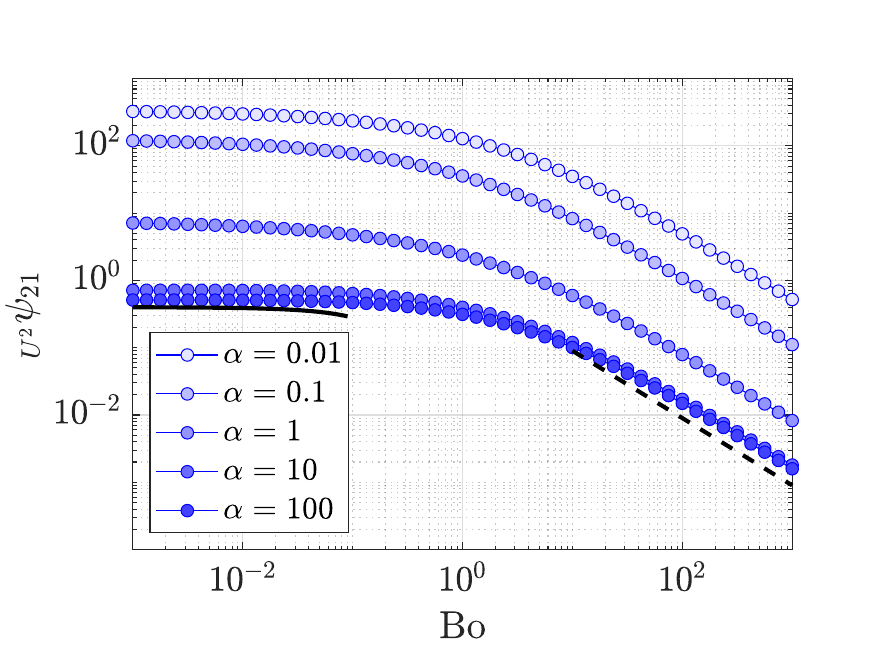}
\caption{Lift coefficient $_{U^2}\psi_{21}$ of the first-order normal force per unit length (see Eq.~(\ref{eq:normal_force})) exerted on the top cylinder as a function of the Bond number Bo and for various gap ratios $\alpha$. This coefficient was obtained from numerical integration of $P_{21}$ along $X$, as computed for $M= 1$, $\beta=99$, $V_1 = 0$, and $V_2 = 1$. The solid line corresponds to an affine decrease with Bo, while the dashed line corresponds to a $\sim1/\textrm{Bo}$ power law.}
\label{fig:psi11_vs_Bo} 
\end{figure}

We saw earlier that increasing Bo reduces the deflection of the interface and in turn the first-order pressures. The variation of the lift coefficient $_{U^2}\psi_{21}$ is thus plotted in Fig.~\ref{fig:psi11_vs_Bo} against Bo and for different values of $\alpha$. For small Bo values, the decrease is affine. For large Bo values, the lift coefficient becomes inversely proportional to Bo. Interestingly, in the latter regime, the curvature term in Eq.~(\ref{eq:Laplace_nondim}) becomes negligible, and the interface response becomes Winkler-like~\cite{salez2015elastohydrodynamics}. However, it is to be remarked  that, in contrast to pure Winkler solids, there are still flows in the bottom layer here.

\begin{figure}
\begin{subfigure}[b]{0.45\textwidth}
\caption{}
\centering
\includegraphics[width=8cm]{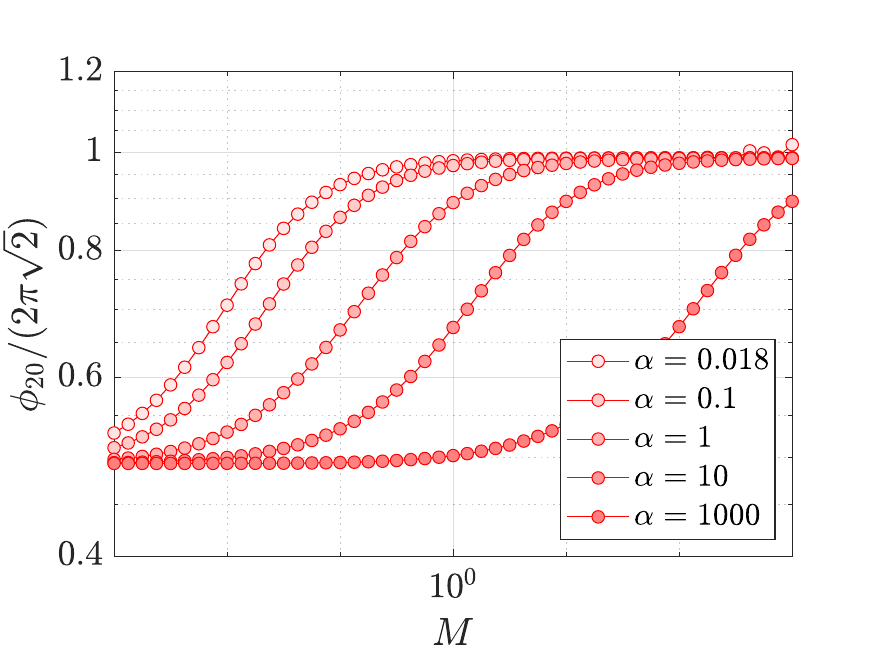}
\end{subfigure}
\hfill
\begin{subfigure}[b]{0.45\textwidth}
\caption{}
\centering
\includegraphics[width=8cm]{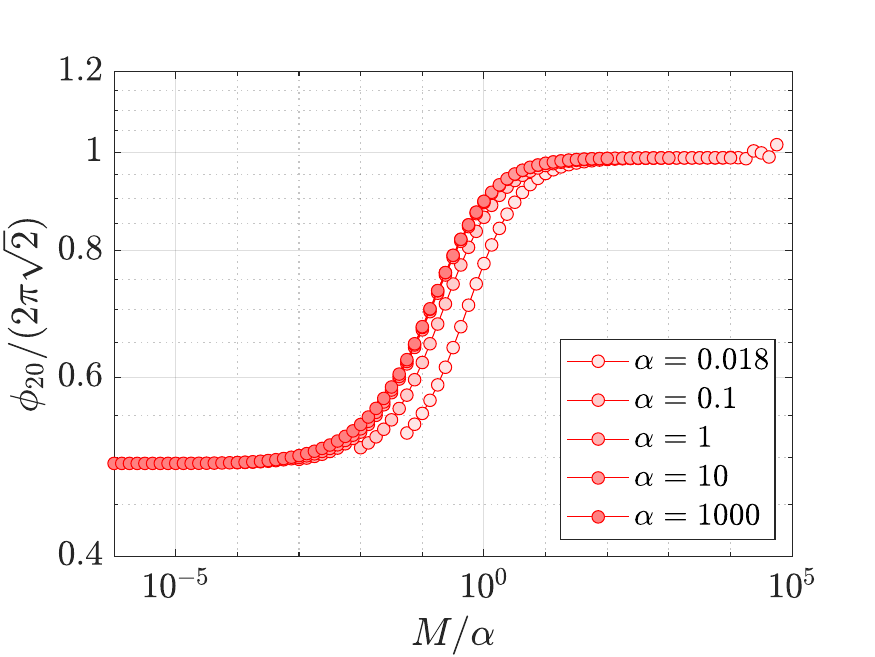} 
\end{subfigure}
\caption{(a) Coefficient $\phi_{20}$ of the zeroth-order torque per unit length (see Eq.~(\ref{eq:torque})) exerted on the top cylinder, and normalized by the value for a rigid no-slip boundary, as a function of the viscosity ratio $M$ and for various gap ratios $\alpha$. This coefficient was computed for $\beta=99$, $V_1 = 0$, and $V_2 = 1$. (b) Same data with a rescaled parameter $M/ \alpha$ on the $x$-axis.}
\label{fig:phi0}
\end{figure}

Finally, in dimensional units, the torque per unit length $T$ generated on the top cylinder is found by integrating the shear stress, as: 
\begin{equation}
T = -a_2\int_{-\infty}^\infty \eta_2\frac{\partial u_2}{\partial z}\bigg\vert_{z=h_2} \textrm{d}x \simeq -\eta_2\omega_2a_2^2\left(\frac{a_2}{d_2}\right)^{1/2}\phi_{20}~,   \end{equation} 
with $\phi_{20} = (2^{3/2}/V_2)\left.\int_{0}^{\infty} \frac{\partial U_{20}}{\partial Z}\right|_{Z = H_2}\textrm{d}X$ the zeroth-order dimensionless coefficient. Indeed, by symmetry of the velocity field, there is no contribution at first order in compliance. The zeroth-order coefficient $\phi_{20}$ is plotted in Fig.~\ref{fig:phi0} against the viscosity ratio $M$, and for different values of the gap ratio $\alpha$. We first see that for all $\alpha$ the zeroth-order coefficient saturates to two limiting values, for $M\rightarrow0$ and $M\rightarrow\infty$ respectively. The latter corresponds to the no-slip rigid case, as expected. Moreover, inspired by our previous study on the normal motion~\cite{jha2023capillary}, we observe a collapse when using the rescaled parameter $M/\alpha$, for $\alpha>1$. 

\section{Conclusion}
We have studied two rotating cylinders separated by a capillary interface in between two lubricating viscous films. Specifically, by using a perturbative expansion in the limit of small deformation of the interface, we have numerically calculated the pressure fields, the interface deflection, and the subsequent force generated on one cylinder. These were separated into lift and inertial-like contributions. We have further investigated the influence of all the relevant geometrical and physical parameters of the problem on these contributions and have revealed a large degree of tunability of their magnitudes and even signs. The latter peculiar feature is absent of classical elastic soft lubrication and highlights the interest of such capillary-lubrication settings. Our results pave the way towards the characterization of colloidal mobility near complex boundaries.

\begin{acknowledgments}
The authors acknowledge financial support from the European Union through the European Research Council under EMetBrown (ERC-CoG-101039103) grant. Views and opinions expressed are however those of the authors only and do not necessarily reflect those of the European Union or the European Research Council. Neither the European Union nor the granting authority can be held responsible for them. The authors also acknowledge financial support from the Agence Nationale de la Recherche under Softer (ANR-21-CE06-0029) and Fricolas (ANR-21-CE06-0039) grants. Finally, they thank the Soft Matter Collaborative Research Unit, Frontier Research Center for Advanced Material and Life Science, Faculty of Advanced Life Science at Hokkaido University, Sapporo, Japan.  
\end{acknowledgments}
\bibliography{Jha2024}
\end{document}